\documentclass[twocolumn,raggedbottom]{aastex701}

\usepackage{amsmath}

\begin{document}

\title{Double White Dwarfs with Circumbinary Disks: Evolution, Population Properties, and Gravitational Wave Signatures}

\author[0000-0002-9343-9797]{Wenwu Ma}
\affiliation{Institute of Theoretical Physics \& Research Center of Gravitation, Lanzhou University, Lanzhou 730000, China}
\affiliation{Lanzhou Center for Theoretical Physics, Key Laboratory of Theoretical Physics of Gansu Province, Key Laboratory of Quantum Theory and Applications of MoE, Gansu Provincial Research Center for Basic Disciplines of Quantum Physics, Lanzhou University, Lanzhou 730000, China}
\email{maww2024@lzu.edu.cn}

\author[0000-0002-1421-4427]{Zhenwei Li}
\affiliation{International Centre of Supernovae (ICESUN), Yunnan  Key Laboratory of Supernova Research, Yunnan Observatories, Chinese Academy of Sciences (CAS), Kunming 650216, People's Republic of China}
\affiliation{University of Chinese Academy of Sciences, Beijing 100049, People's Republic of China}
\email[show]{lizw@ynao.ac.cn}

\author[0009-0006-1130-9777]{Dylan Hebrail}
\affiliation{Lund Observatory, Division of Astrophysics, Department of Physics, Lund University, Box 118, 22100 Lund, Sweden}
\email{dylan.hebrail@fysik.lu.se}

\author[0000-0003-0821-4583]{Dandan Wei}
\affiliation{Institute of Science and Technology Austria: Klosterneuburg, Lower Austria, AT}
\email{dandan.wei@hotmail.com}

\author[0000-0002-1421-4427]{Hongwei Ge}
\affiliation{International Centre of Supernovae (ICESUN), Yunnan  Key Laboratory of Supernova Research, Yunnan Observatories, Chinese Academy of Sciences (CAS), Kunming 650216, People's Republic of China}
\affiliation{University of Chinese Academy of Sciences, Beijing 100049, People's Republic of China}
\email{gehw@ynao.ac.cn}

\author[0000-0001-9204-0779]{Ross Church}
\affiliation{Lund Observatory, Division of Astrophysics, Department of Physics, Lund University, Box 118, 22100 Lund, Sweden}
\email{ross.church@fysik.lu.se}

\author[0000-0001-5284-8001]{Xuefei Chen}
\affiliation{International Centre of Supernovae (ICESUN), Yunnan  Key Laboratory of Supernova Research, Yunnan Observatories, Chinese Academy of Sciences (CAS), Kunming 650216, People's Republic of China}
\affiliation{University of Chinese Academy of Sciences, Beijing 100049, People's Republic of China}
\email[show]{cxf@ynao.ac.cn}

\author[0000-0002-4117-4176]{Yuxiao Liu}
\affiliation{Institute of Theoretical Physics \& Research Center of Gravitation, Lanzhou University, Lanzhou 730000, China}
\affiliation{Lanzhou Center for Theoretical Physics, Key Laboratory of Theoretical Physics of Gansu Province, Key Laboratory of Quantum Theory and Applications of MoE, Gansu Provincial Research Center for Basic Disciplines of Quantum Physics, Lanzhou University, Lanzhou 730000, China}
\email[show]{liuyx@lzu.edu.cn}

\begin{abstract}

Double white dwarfs (DWDs) are important multi-messenger sources and powerful probes of binary interaction physics. Close DWDs formed through isolated binary evolution are expected to have negligible eccentricities. However, recent observations have revealed that some eclipsing DWDs exhibit eccentricities of order $10^{-3}$ or larger. In this work, we investigate the interaction between a circumbinary disk (CBD) and the central DWD, where the CBD is assumed to form from a recent common-envelope (CE) ejection. Our models are the first to include CBD–DWD interaction in DWD population synthesis and to constrain disk parameters such as disk mass and lifetime. Our population models predict a correlation between the eccentricity distribution and the orbital period distribution of DWDs, which can be tested by future observations. 
Furthermore, we discuss the influence of CBD on the gravitational-wave (GW) signal. The CBD-DWD interaction can shrink the orbit and excite eccentricity, extending the time for which the system is individually resolvable by a future space-borne GW detector. Our results suggest that CBD interaction should be accounted for in population synthesis of Galactic DWD GW sources and the stochastic GW foreground.
\end{abstract}


\keywords{
  \uat{Binary stars}{154} ---
  \uat{White dwarf stars}{1799} ---
  \uat{Circumstellar disks}{235} ---
  \uat{Common envelope evolution}{2154} ---
  \uat{Gravitational waves}{678}
}

\section{Introduction}
\label{sec:intro}

Double white dwarf (DWD) binaries are among the most numerous
populations of compact remnant binaries in the Galaxy, with a
theoretically predicted population of $\sim 10^8$
systems \citep[e.g.,][]{Nelemans2001a368,Nelemans2001b,Nelemans2001a365,Ruiter2010,YuJeffery2010,Nissanke2012,Korol2017,Maoz2018DWD,Korol2019,Lambertslisa,Breivik2020DWDforeground,Huang2020TianQinDWD,LiZW2020,Korol2022DWD,2023A&A...669A..82L}.
Double white dwarfs (DWDs) with orbital periods shorter than $\sim1$ hr emit gravitational wave (GW) at millihertz frequencies that
will be directly accessible to the future space-borne GW antenna, such as Laser Interferometer Space Antenna
(LISA; \citealt{Amaro-Seoane2023}), TaiJi (\citealt{HuWu2017Taiji}), TianQin (\citealt{Luo2016TianQin}), among which tens of thousands of DWDs are expected to be individually resolved \citep[e.g.,][]{Ruiter2010,YuJeffery2010,Nissanke2012,Korol2017,Lambertslisa,LiZW2020,Zhang2021GBSIEVER,Korol2022DWD,Zhang2022Network,2023A&A...669A..82L,Kupfer2024,Zhang2025GBSIEVERC,Zhao2026DWDResolutionPrior,Zhao2026MilkyWayDWD}. As GW emission removes orbital energy, short-period DWDs spiral inward on timescales of several hundred Myr to several Gyr, may eventually lead to mass transfer or coalescence \citep{1984ApJ...277..355W,2002MNRAS.329..897H}, which then may produce AM CVn systems, R~Coronae Borealis stars, and possibly a subclass of Type~Ia supernovae \citep{1984ApJ...277..355W,RuiterSeitenzahl2025}. 

DWDs are generally thought to form through two episodes of mass transfer in a primordial binary system. The initially more massive primary evolves first, expands, and fills its Roche lobe. Depending on the initial mass ratio and orbital separation, this first mass transfer episode may proceed either as dynamically stable Roche-lobe overflow (RLOF) or as a common-envelope (CE) event. In either case, the hydrogen-rich envelope of the primary is removed, leaving a bare core that cools to become the first white dwarf (WD) \citep{1984ApJ...277..355W,2013A&ARv..21...59I}. The secondary then evolves, expands, and fills its Roche lobe while its companion is already a compact WD, this second mass transfer phase may likewise be either stable or unstable. In principle, such binary interaction processes would erase most of the orbital eccentricity through tidal dissipation during RLOF and through dynamical friction during CE evolution \citep{Paczynski1976,Zahn1977,2002MNRAS.329..897H,2013A&ARv..21...59I}. Consequently, DWDs formed through isolated binary evolution are expected to have negligible eccentricities \citep{Willems2007EccentricDWD,Kaplan2010EclipseTiming}. However, recent time-domain photometric surveys, in particular the Zwicky Transient Facility (ZTF; \citet{Burdge2019,Burdge2020,2025arXiv250515580V}), have reported secondary-eclipse timing constraints at the projected scale $|e\cos\omega|\sim10^{-3}$ for eclipsing DWDs, where $\omega$ is the longitude of periastron. This is several orders of magnitude higher than the essentially negligible residual eccentricity expected from these binary interaction processes \citep{Zahn1977,1992RSPTA.341...39P}. While some studies have proposed that mass transfer processes may retain a small residual eccentricity (e.g., \citealt{Rocha2025}), such scenarios have not yet been broadly incorporated into standard binary evolution models, and whether they can account for the observed eccentricities remains an open question. 

The resonant interaction between a circumbinary disk (CBD) and the central binary is a promising mechanism to excite the eccentricity of the inner binary: linear perturbation theory shows that outer Lindblad resonances transfer orbital angular momentum to the disk and can excite the binary eccentricity \citep{1979ApJ...233..857G,1991ApJ...370L..35A,1994ApJ...421..651A,1996LNP...465..115A}. This mechanism has been applied to post mass transfer and post-asymptotic giant branch binaries and explored in recent CBD evolution models \citep{2013A&A...551A..50D,2014ApJ...797L..24A,Rafikov2016,2023MNRAS.521...35I,2024A&A...688A.128V}. During CE ejection, a fraction of the envelope may fail to reach escape velocity and remain gravitationally bound to the binary after the dynamical inspiral phase \citep{2011MNRAS.417.1466K,2016MNRAS.461..486K,2013A&ARv..21...59I}. Three-dimensional simulations show that this residual gas can settle into a rotationally supported, disk-like structure whose mass and angular momentum transport can modify the subsequent orbital evolution \citep{2023A&A...674A.121G,2024A&A...691A.244V}. The subsequent evolution of post-CE CBDs and their effects on compact-binary orbits and final outcomes have also been explored \citep{2023ApJ...955..125T,2024A&A...688A..87W,2026ApJ...998...79U}, for stellar-mass binary black holes, a prior CBD-driven phase may leave a LISA-band eccentricity signature \citep{RomeroShaw2024CBDLISA}. Motivated by the recently reported projected eccentricity constraints in eclipsing DWDs, we investigate whether CBD--DWD interaction can explain these observations, model the population properties of Galactic DWDs with CBDs, and constrain the corresponding disk parameters. We further propose that CBD interaction can affect the GW radiation properties of DWDs in extremely short orbits, potentially leaving detectable signatures for future space-borne GW detectors.

The remainder of this paper is organized as follows. Section~\ref{sec:Methods} describes the CBD--DWD interaction model, including the resonant torque, disk accretion, and GW-driven orbital evolution, and examines the resulting eccentricity evolution of DWDs interacting with CBDs. Section~\ref{sec:bps} describes the binary population synthesis (BPS) model and the resulting population properties. Section~\ref{sec:lisa} discusses the implications for future GW detection. Section~\ref{sec:conclusions} summarizes the main conclusions.

\section{CBD-DWD interaction} 
\label{sec:Methods}
\subsection{CBD Models}
\label{sec:cbd}
Once the second WD is produced from the successful ejection of CE, we assume that part of material will be remain gravitationally bound and settle into a CBD \citep{2011MNRAS.417.1466K,2016MNRAS.461..486K,2024A&A...691A.244V}. In this work, we investigate two key aspects of the interaction between the binary and the CBD: (1) resonant interaction between the CBD and the binary, (2) mass accretion from the CBD onto the binary. We note that these two processes are treated simultaneously in our model. Both channels contribute to the angular-momentum budget: accretion carries the specific angular momentum of the disk material at the inner edge into the binary, while resonant torques exchange angular momentum between the binary and the disk. In addition, the GW emission of the inner binary is always included in the simulations. The details of our methods are introduced as follows.

\noindent\textbf{(1) Resonant Interaction:} We model the CBD--DWD coupling using linear perturbation theory
\citep{1979ApJ...233..857G,1994ApJ...421..651A,1996ASIC..477...53L,1996LNP...465..115A}
and its application to post-CE binaries
\citep{2013A&A...551A..50D,2014ApJ...797L..24A}.
We expand the binary gravitational potential in Fourier harmonics with
integer indices and pattern speeds. 
These components excite density waves at Lindblad and
corotation resonances.
For density-independent kinematic viscosity, we estimate the resonant
torque by balancing the rate of angular momentum transfer to the disk
against the viscous torque \citep{2013A&A...551A..50D}.
The semimajor-axis evolution is
\citep{1996LNP...465..115A,2013A&A...551A..50D,2014ApJ...797L..24A}
\begin{equation}
    \left(\frac{\dot{a}}{a}\right)_{\rm res}
    = -\frac{2l}{m}\frac{M_{\rm disk}}{\mu}\,\alpha
      \left(\frac{H}{R}\right)^{\!2}
      \frac{a}{R}\,\Omega_{\rm b},
    \label{eq:adot_res}
\end{equation}
where $l$ and $m$ are integer harmonic indices,
  $\mu=M_1M_2/(M_1+M_2)$ is the reduced mass, $R$ is the CBD
  half-angular-momentum radius, $H/R$ is the aspect ratio,  $M_{\rm disk}$ is the
   CBD mass, $a$ is semimajor axis, $\Omega_{\rm b}$ is orbital angular velocity, and $\alpha$ is the viscosity parameter. We adopt $l=1$, $m=2$, $\alpha=0.1$ and $H/R=0.1$ \citep{1996LNP...465..115A,2013A&A...551A..50D,2014ApJ...797L..24A}. We set the half-angular-momentum radius to $R(t)=250a(t)$ in the torque prescription of \citet{2013A&A...551A..50D}. This is a simplified version of the estimate $R=\sqrt{r_{\rm in}r_{\rm out}}$ adopted by \citet{2013A&A...551A..50D}, where $r_{\rm in}$ and $r_{\rm out}$ are the disk inner and outer radii. We emphasize that $R$ is not computed self-consistently here. In a real CBD, $R$ evolves with the surface-density profile, viscous spreading,and mass loss, and $R/a$ may itself depend on the binary separation, disk mass, and evolutionary stage \citep{2023ApJ...955..125T}. As a consistency check, we verified that, for the disk parameters adopted by \citet{2014ApJ...797L..24A}, this choice reproduces the characteristic eccentricity evolution obtained in that study.
  
The CBD mass, $M_{\rm disk}$, depends on the retained envelope mass and the adopted disk prescription \citep{2011MNRAS.417.1466K,2016MNRAS.461..486K,2023ApJ...955..125T,2026ApJ...998...79U}. We treat its initial value as a free parameter, varying it from \(10^{-5}\) to several \(10^{-3}\,M_\odot\), and explore its effect on the results. We assume a constant average mass-loss rate for the CBDs, given by \(\dot{M}_{\rm disk} = -M_{\rm disk,0}/\tau_{\rm disk}\) \citep{2024A&A...688A..87W}, where \(\tau_{\rm disk}\) is the disk lifetime, which spans a wide range of \(\sim 10^2\)--\(10^5\ \rm yr\). We first treat the disk lifetime as a free parameter and discuss its effect. The lifetime models adopted in our binary population synthesis model are described in Section~\ref{sec:bps_input}.

For eccentricity growth, we adopt Eq.~(15) of \citet{2024A&A...688A..87W}.
It joins the low-eccentricity expression
\citep{1996LNP...465..115A,2013A&A...551A..50D,2014ApJ...797L..24A} to a
fit to the same smoothed particle hydrodynamics (SPH) results at higher eccentricity:
\begin{equation}
  \dot{e}_{\rm res} =
  \left\{
  \begin{array}{@{}l@{}}
  \displaystyle
  \dfrac{2(1-e^2)}{e + \frac{\alpha}{100e}}
  \left(\dfrac{l}{m} - \dfrac{1}{\sqrt{1-e^2}}\right)
  \left(\dfrac{\dot{a}}{a}\right)_{\rm res},
  \quad e \leq 0.2, \\[4mm]
  \displaystyle
  \left(\dfrac{C_1}{e} + C_2\right)
  \left(\dfrac{\dot{a}}{a}\right)_{\rm res},
  \quad 0.2 < e \leq e_{\rm res,max}, \\[3mm]
  0, \quad e > e_{\rm res,max}.
  \end{array}
  \right.
  \label{eq:edot_res}
  \end{equation}
with $C_1=-1.3652$, $C_2=1.9504$ and $e_{\rm res,max}=0.7$.
The resonant torque transfers angular momentum from the binary to the
disk, shrinking the orbit and exciting eccentricity.

\noindent\textbf{(2) Orbital Evolution Driven by Disk Accretion:}\label{sec:acc} 
Gas within the CBD transports its angular momentum outward. As a consequence, gas that loses angular momentum migrates inward toward the inner edge of the CBD and may ultimately be accreted onto the binary \citep{2023ARA&A..61..517L}. Following \citet{2024A&A...688A..87W}, we assume that the entire mass lost from the CBD is accreted by the inner binary, with an accretion rate of \(\dot{M}_{\rm acc} \equiv -\dot{M}_{\rm disk}\).
The gas is shared unequally between the primary ($M_1$) and secondary
($M_2$) according to the preferential-accretion ratio
$\lambda(q,e)\equiv\dot{M}_1/\dot{M}_2$, with $q=M_2/M_1\leq1$.
We obtain $\lambda$ by bilinear interpolation in $(q,e)$ on a grid
calibrated on hydrodynamical simulations and hold it constant outside
the tabulated range \citep{2023MNRAS.522.2707S,2023MNRAS.518.5059S}.
Mass conservation gives $\dot{M}_{\rm acc} = \dot{M}_1+\dot{M}_2$.

We assume that gas arrives from the cavity edge at $r_{\rm in}=2a$
\citep{1994ApJ...421..651A} with specific angular momentum
$\ell_{\rm acc}=\sqrt{2\,G M a}$, where $M=M_1+M_2$.
The angular momentum supply rate is
$\dot{J}_{\rm acc}=\ell_{\rm acc}\,\dot{M}_{\rm acc}$.
Using $J_b=(M_1M_2/M)\sqrt{G M a}$ for the binary orbital angular
momentum, the accretion-driven semimajor-axis evolution is
  \begin{equation}
    \left(\frac{\dot{a}}{a}\right)_{\rm acc}
    = \frac{\dot{M}}{M}
    + 2\,\frac{\dot{J}_{\rm acc}}{J_b}
    - 2\,\frac{\dot{M}_1}{M_1}
    - 2\,\frac{\dot{M}_2}{M_2}\,.
    \label{eq:adot_res_acc}
  \end{equation}
We add $\dot{a}_{\rm acc}$ to the resonant and GW terms in
Equation~(\ref{eq:adot_total}) and switch it off after disk exhaustion.
We neglect direct eccentricity changes from accretion
\citep{2013A&A...551A..50D}. The effect of accretion on the orbital evolution is minor compared with that of the resonant torque. Switching off the accretion terms changes the final eccentricities by less than $1\%$ in most cases, and by at most $\sim10\%$ for the most massive and short-lived disks. Previous CBD models of eccentricity pumping neglected accretion \citep{2013A&A...551A..50D,2014ApJ...797L..24A}.

\noindent\textbf{(3) GW Radiation:} GW emission is included throughout the evolution. The semimajor-axis decay rate $\dot{a}_{\rm GW}$ and eccentricity damping rate $\dot{e}_{\rm GW}$ follow the orbit-averaged quadrupole expressions of \citet{1963PhRv..131..435P,1964PhRv..136.1224P}.

The combined orbital evolution is
  \begin{align}
      \dot{a} &= \dot{a}_{\mathrm{res}} + \dot{a}_{\mathrm{GW}} + \dot{a}_{\mathrm{acc}},
      \label{eq:adot_total}\\
      \dot{e} &= \dot{e}_{\mathrm{res}} + \dot{e}_{\mathrm{GW}}.
      \label{eq:edot_total}
  \end{align}
These equations are integrated with the \texttt{DOP853} Runge--Kutta solver \citep{HairerNorsettWanner1993}, using relative and absolute tolerances of $10^{-8}$ and $10^{-11}$, respectively. During the disk phase, while $t\leq\tau_{\rm disk}$ and $M_{\rm disk} > 0$, the resonant semimajor-axis
and accretion terms remain active. After disk dispersal or exhaustion, all disk terms vanish and the subsequent
evolution is driven by GW emission alone. Once one of the WD components fills its Roche lobe, the binary is assumed to merge, with the zero-temperature WD radius from \citet{1972ApJ...175..417N} adopted for this criterion.

We set the initial binary eccentricity before the CBD formation according to the equilibrium relation of \citet{1992RSPTA.341...39P},
\begin{equation}
  e_{\rm eq}=1.5\times10^{-4}\times\left(\frac{P_{\rm orb}}{100\,{\rm d}}\right),
  \label{eq:initial_ecc}
\end{equation}
where $P_{\rm orb}$ is the initial orbital period.

\subsection{Eccentricity evolution with CBD}
\label{sec:ecc_or_p}

\begin{figure}[ht!]
 \centering
 \includegraphics[width=\columnwidth]{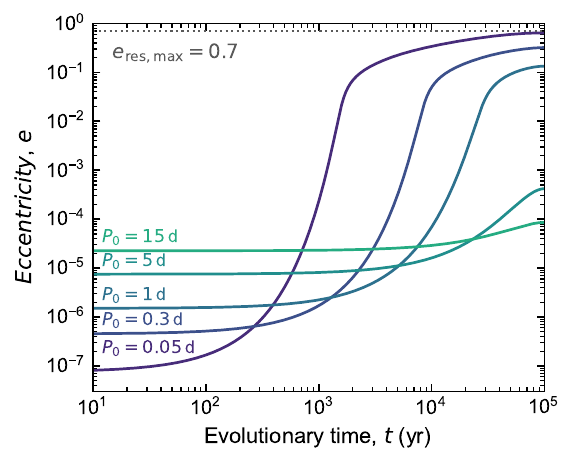}
 \caption{Time evolution of the eccentricity $e$ for different initial orbital
 periods $P_0$. The DWDs have component masses $M_1=1.25\,M_\odot$ and $M_2=0.28\,M_\odot$. The evolutionary time $t$ is measured from the start of the
 disk interaction and spans the disk lifetime. The initial disk mass and lifetime
 are fixed at $M_{\rm disk,0}=10^{-5}\,M_\odot$ and
 $\tau_{\rm disk}=10^{5}\,{\rm yr}$. Colored
 curves show five representative models, $P_0=0.05$, $0.3$, $1$, $5$, and
 $15\,{\rm d}$, from dark purple (short periods) to green (long periods). Each
 curve starts from its period-dependent initial eccentricity given by
 Equation~(\ref{eq:initial_ecc}). The dotted gray
 line marks the resonance limit $e_{\rm res,max}=0.7$. Short-period binaries have higher resonant growth rates and are excited first, whereas long-period binaries show smaller eccentricity increases during the disk lifetime.
 \label{fig:evol}}
 \end{figure}

Figure~\ref{fig:evol} shows several representative examples of the eccentricity evolution at a fixed disk lifetime. The DWDs have component masses $M_1=1.25\,M_\odot$ and $M_2=0.28\,M_\odot$, with orbital periods ranging from $0.05$ to $15\,{\rm d}$. The initial disk mass is fixed at $M_{\rm disk,0}=10^{-5}\,M_\odot$ and the disk lifetime at $\tau_{\rm disk}=10^{5}\,{\rm yr}$. For different disk masses and lifetimes, the rate of eccentricity change varies, but the overall trend of the eccentricity evolution remains qualitatively unchanged. The initial eccentricity follows Equation~(\ref{eq:initial_ecc}). Resonant interactions gradually increase the eccentricity. At fixed binary and disk masses, eccentricity, and dimensionless disk parameters, Equation~(\ref{eq:edot_res}) gives $\dot{e}_{\rm res}\propto P_{\rm orb}^{-1}$ through $\Omega_{\rm b}=2\pi/P_{\rm orb}$, so short-period systems experience stronger excitation. For example, the $P_0=0.05$ and $0.3\,{\rm d}$ models reach $e\simeq0.64$ and $0.32$, respectively, during the disk interaction, while the $P_0=1\,{\rm d}$ model reaches $e\simeq0.13$. The growth is weaker at longer periods: for $P_0=15\,{\rm d}$, the eccentricity increases from $2.25\times10^{-5}$ to about $8.6\times10^{-5}$ and remains small. This resonant excitation is consistent with the CBD interaction calculations of \citet{2014ApJ...797L..24A} for neutron star (NS)--He WD binaries.

\begin{figure*}[!t]
 \centering
 \includegraphics[width=\textwidth]{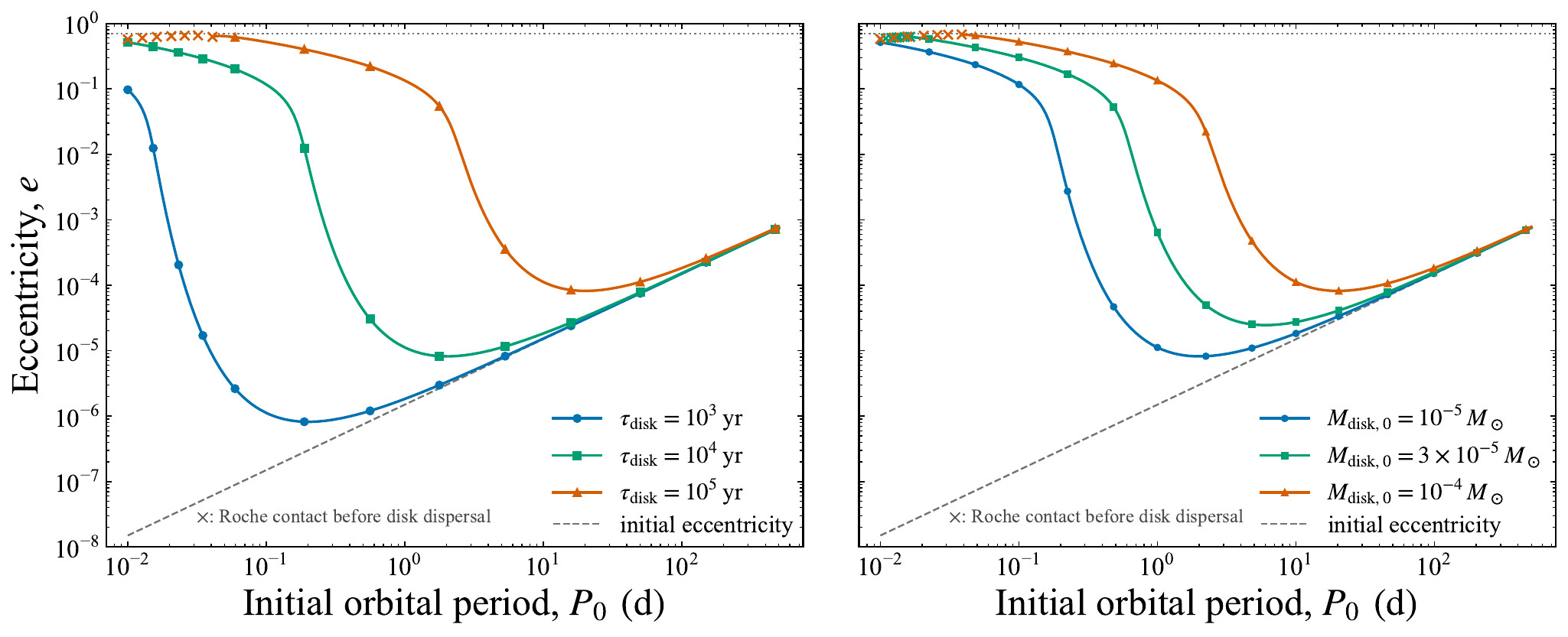}
 \caption{The final eccentricity after the disk–binary interaction as a function of the initial orbital period. The DWDs have component masses $M_1=1.25\,M_\odot$ and $M_2=0.28\,M_\odot$.  Left: CBD mass is fixed at $M_{\rm disk,0}=10^{-5}\,M_\odot$, and three cases of disk lifetime, $\tau_{\rm disk}=10^3$, $10^4$, and $10^5\,{\rm yr}$, are considered. Right: results for
  $M_{\rm disk,0}=10^{-5}$, $3\times10^{-5}$, and $10^{-4}\,M_\odot$
  with $\tau_{\rm disk}=10^4\,{\rm yr}$. The curves show eccentricities
  at disk dispersal for surviving systems. The crosses show eccentricities
  at each system's Roche-contact time before disk dispersal. The dashed
  gray lines show the initial eccentricity from Equation~(\ref{eq:initial_ecc}).
 \label{fig:summary}}
 \end{figure*}

Figure~\ref{fig:summary} illustrates how the eccentricity excitation depends on the CBD lifetime and initial CBD mass. In the left panel, the CBD mass is fixed at $M_{\rm disk,0}=10^{-5}\,M_\odot$, and three disk lifetimes are considered: $\tau_{\rm disk}=10^3$, $10^4$, and $10^5\,{\rm yr}$. In the right panel, the disk lifetime is fixed at $\tau_{\rm disk}=10^4\,{\rm yr}$, and three disk masses are considered: $M_{\rm disk,0}=10^{-5}$, $3\times10^{-5}$, and $10^{-4}\,M_\odot$. For a binary with all other initial parameters fixed, a longer disk lifetime or a larger disk mass generally results in a larger eccentricity following the disk interaction. For short orbital periods, the eccentricity can be excited to a large value, whereas for long periods it remains close to its initial value, as shown in Figure~\ref{fig:evol}. Orbital shrinkage and eccentricity growth can bring very short-period binaries into Roche contact during the disk interaction \citep{1983ApJ...268..368E, 2007ApJ...660.1624S}. We assume that the binary merges once one of the WD components fills its Roche lobe. Systems that reach contact during the disk phase are classified as disk-driven mergers. Although we do not follow the subsequent evolution in this case, disk-driven mergers have important implications for GW signatures, which will be explored in Section~4.

\section{binary population synthesis}
\label{sec:bps}

\subsection{Main inputs}
\label{sec:bps_input}
In this work, we assume that if the second mass transfer phase is unstable and enters a CE phase, a CBD forms from the bound fallback material surrounding the newly born DWD and subsequently affects its orbital evolution. Accordingly, our simulations consist of two parts: (i) the formation of the DWD immediately after CE ejection, modeled via BPS, and (ii) the post-formation interaction between the CBD and the DWD, modeled with analytical prescriptions for resonant torques and accretion. The main inputs of our BPS models are introduced as follows.

We simulate the formation of DWDs using the rapid Binary Star Evolution (\texttt{BSE}) code \citep{2002MNRAS.329..897H}, incorporating the updates described in \citet{2023A&A...669A..82L}. The input parameters and physical prescriptions are summarized as follows. During binary evolution, we include GW radiation, magnetic braking, wind accretion, tidal friction, stable RLOF, and CE evolution. The first four processes are treated with the default parameters of \texttt{BSE}. For stable RLOF, accretion is limited by the thermal-timescale mass-transfer rate of the accretor, assuming a maximum accretion efficiency of 0.5. For CE evolution, we set the CE ejection efficiency to $\alpha_{\rm CE} = 1.0$. The critical mass ratio that determines the stability of mass transfer is adopted from the evolutionary grids of \citet{2010ApJ...717..724G, 2015ApJ...812...40G, 2020ApJ...899..132G}. For the population synthesis, we adopt a Salpeter initial mass function for the primary star over the mass range $0.1$--$100\,M_\odot$ \citep{2023A&A...669A..82L}. The initial mass ratio $q = M_2/M_1$ is drawn from a uniform distribution between 0 and 1, and the initial orbital separation is sampled following the distribution of \citet{1998MNRAS.296.1019H}. The Galactic star formation rate is assumed to be $5\,M_\odot\,{\rm yr}^{-1}$ over a Hubble time of 13.8 Gyr. The above input parameters are broadly consistent with the fiducial model of \citet{2023A&A...669A..82L}, which has been shown to reproduce the observed DWD populations well. We restrict the initial orbital periods to $P_0\leq100$ days, as longer-period systems are not relevant for the formation channel considered here.

In this work, the disk lifetime is defined by
\begin{equation}
\tau_{\rm disk}=\frac{M_{\rm disk,0}}{\dot M_{\rm acc}},
\qquad \dot M_{\rm acc}=-\dot M_{\rm disk},
\end{equation}
all mass lost from the disk is assumed to be accreted by the inner binary. Since
the disk loses mass only through accretion, $\dot M_{\rm acc}$ is also
the rate at which gas is transported inward by viscosity. The disk
lifetime is therefore expected to be of order the viscous time at the
characteristic disk radius,
\begin{equation}
\tau_{\rm disk}\sim t_\nu\sim \frac{R^2}{\nu},
\end{equation}
where $\nu$ is the kinematic viscosity (e.g.,
\citealt{2013A&A...551A..50D,2023ARA&A..61..517L,2023ApJ...955..125T}). For an $\alpha$ disk with constant $\alpha$ and
$H/R$,
\begin{equation}
\nu=\alpha c_{\rm s}H
=\alpha\left(\frac{H}{R}\right)^2R^2\Omega
\propto R^{1/2},
\end{equation}
since $\Omega\propto R^{-3/2}$ for a Keplerian disk. Hence
$t_\nu\propto R^{3/2}$. With $R\propto a$, as assumed in Section~\ref{sec:cbd},
this gives $\tau_{\rm disk}\propto a^{3/2}$.

In reality, additional processes such as photoevaporation and
magnetically driven winds can modify this scaling \citep{2023ApJ...955..125T}, and the normalization is not fixed by the viscous argument alone. We therefore parameterize
the disk lifetime as
\begin{equation}
\tau_{\rm disk}=C a^{\gamma}.
\end{equation}
For each $\gamma$, the coefficient $C$ is fixed by requiring a maximum disk lifetime of $10^5$~yr at the orbital-period cutoff of $100$~d adopted above \citep[e.g.,][]{2013A&A...551A..50D,2014ApJ...797L..24A}. We
consider three cases: $\gamma=1$ (Model~A), $\gamma=3/2$ (Model~B),
which is the viscous scaling for constant $H/R$, and $\gamma=2$
(Model~C), so that Models~A and~C bracket the viscous scaling.

\subsection{Population properties of DWDs with CBDs}
\label{sec:bps_pe}

\begin{figure}[!t]
  \centering
  \includegraphics[width=\columnwidth]{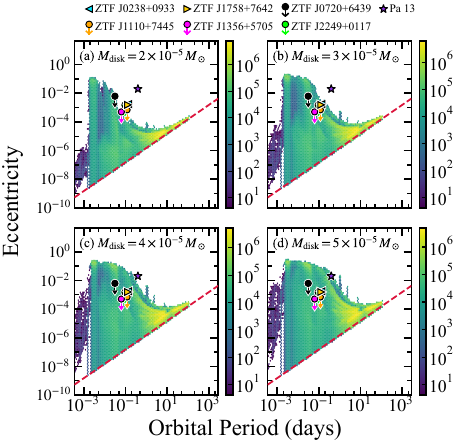}
  \vspace{-3pt}
  \caption{Present-day orbital period--eccentricity distributions for the
lifetime prescription of Model~A. Panels (a)--(d) use $M_{\rm disk,0}=(2,3,4,5)\times10^{-5}\,M_\odot$, respectively. The total number of DWDs varies among the four panels due to disk-driven merger events, but all values are close to \(4.37\times10^8\), differing by only on the order of \(10^5\). Systems driven to Roche contact by CBD interaction are treated as mergers and are therefore excluded from the counts. The color scale gives the
  number of Galactic DWDs per bin. The red dashed line is the adopted
  \citet{1992RSPTA.341...39P} relation. Downward arrows indicate upper limits on the eccentricity constraints, all
  other symbols denote measured eccentricities
  \citep{2025arXiv250515580V,Reindl2026Pa13}.}
  \label{fig:pe_modela}
\end{figure}

\begin{figure}[!t]
  \centering
  \includegraphics[width=\columnwidth]{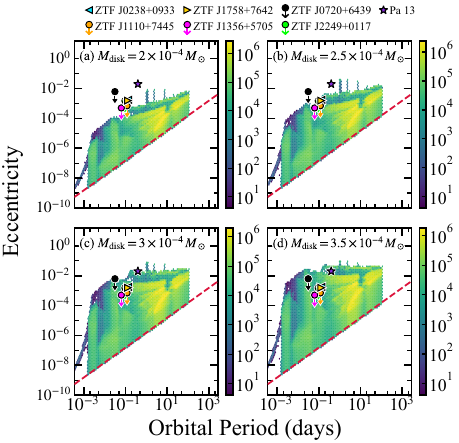}
  \vspace{-3pt}
  \caption{Similar as Figure~\ref{fig:pe_modela}, but for the viscous lifetime scaling of Model~B, $\tau_{\rm disk}\propto a^{3/2}$. Panels (a)--(d) use $M_{\rm disk,0}=(2,2.5,3,3.5)\times10^{-4}\,M_\odot$, respectively. The total number of DWDs is close to \(4.37\times10^8\) in all panels, differing by only on the order of \(10^4\).}
  \label{fig:pe_modelb}
\end{figure}

\begin{figure}[!t]
  \centering
  \includegraphics[width=\columnwidth]{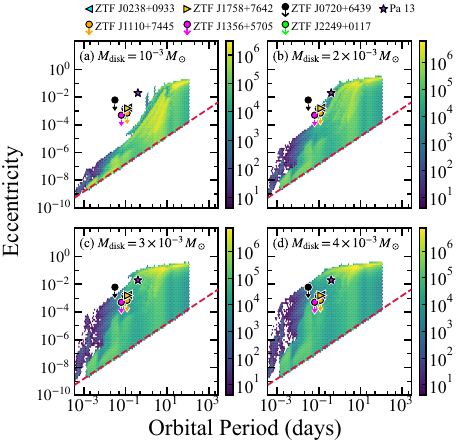}
  \vspace{-3pt}
  \caption{Similar as Figure~\ref{fig:pe_modela}, but for the lifetime prescription of Model~C, $\tau_{\rm disk}\propto a^{2}$. Panels (a)--(d) use $M_{\rm disk,0}=(1,2,3,4)\times10^{-3}\,M_\odot$, respectively. The total number of DWDs varies among the four panels due to disk-driven merger events, but all values are close to \(4.44\times10^8\), differing by only on the order of \(10^5\).}
  \label{fig:pe_modelc}
\end{figure}

The primary effect of the CBD--DWD interaction is on the eccentricities and orbital periods of the binaries, while its effect on the DWD mass distribution is negligible. We therefore focus on the distributions of eccentricity and orbital period. Figures~\ref{fig:pe_modela}, \ref{fig:pe_modelb}, and~\ref{fig:pe_modelc} show the present-day distributions of DWDs for Models~A, B, and~C, respectively. All models adopt the \citet{1992RSPTA.341...39P} relation to set the initial eccentricities of all DWD systems. For comparison, we also show the seven eccentric DWDs from observations: the six ZTF systems with secondary-eclipse timing constraints on \(e\cos\omega\) \citep{2025arXiv250515580V}, together with Pa~13, for which the true \(e\) is estimated from model-based light-curve and radial-velocity fitting \citep{Reindl2026Pa13}. Note that Pa~13 is a binary of twin hot pre-WDs inside a planetary nebula, which may have been produced by a recent CE ejection event.

Based on the binary population synthesis inputs and the disk models, we obtain total DWD numbers in the present-day Galaxy of \(\sim 4.37\times10^8\), \(\sim 4.37\times10^8\), and \(\sim 4.44\times10^8\) for Models~A, B, and~C, respectively. The difference in disk mass and lifetime affects the total number of DWDs through disk-driven merger events. The three disk lifetime models also yield distinct eccentricity distributions for DWDs, as shown in Figures~\ref{fig:pe_modela}--\ref{fig:pe_modelc}. In Model~A, the disk lifetime is assumed to be proportional to the binary separation, in Model~B, to $a^{3/2}$, whereas in Model~C, it is proportional to the square of the binary separation. At a fixed initial period, the disk lifetime is longest in Model~A and shortest in Model~C, with Model~B in between. Consequently, in Model~A, even a relatively small disk mass can excite the orbit to a large eccentricity. In Model~B, the disk lifetime is intermediate, so an intermediate disk mass is required. In contrast, in Model~C, the disk lifetime is relatively short for short-period systems, so a larger disk mass is required to match the observed sample. In Figure~\ref{fig:pe_modela}, the stronger excitation toward shorter periods in Model~A reflects the competition between stronger resonant forcing and shorter disk lifetimes. In Figure~\ref{fig:pe_modelb}, Model~B shows an intermediate behavior: the whole distribution moves to higher eccentricities as the disk mass increases.
In Figure~\ref{fig:pe_modelc}, however, because short-period systems have shorter disk lifetimes, their eccentricities cannot be excited to large values during the disk--binary interaction, whereas the more massive disks allow long-period systems to develop large eccentricities.

The eccentricity of a DWD is difficult to measure directly. Eclipse timing is one of the most important methods for detecting eccentricity signals, and numerous eclipsing DWDs have been discovered (e.g., \citealt{2014ApJ...792...39H,Kaplan2014NLTT11748,Burdge2020ZTFSearch,2025arXiv250515580V,Reindl2026Pa13}).
For the six ZTF systems considered here, secondary-eclipse timing constrains \(e\cos\omega\), yielding two nonzero detections and four upper limits \citep{2025arXiv250515580V}. Because eclipses are more likely to occur in short-period systems, eclipse-based eccentricity constraints are essentially unavailable for wide-orbit DWDs. Consequently, such systems are often assumed to be circular in observational analyses. In our population models, an important distinction among the different disk-lifetime prescriptions is the relation between eccentricity and orbital period. Model~A predicts that long-period systems (\(P_{\rm orb}\gtrsim1\,{\rm d}\)) typically have small eccentricities (\(e\sim10^{-6}\)--\(10^{-4}\)), Model~B predicts intermediate eccentricities (\(e\sim10^{-4}\)--\(10^{-2}\)) and Model~C predicts that long-period systems can retain considerable eccentricities (\(e\gtrsim0.1\)). Therefore, future observations of eccentric long-period DWDs, or robust upper limits on their absence, would place strong constraints on the disk models, although such measurements remain observationally challenging. It should be noted that Pa~13 has been proposed to be a recent CE ejection event, and \citet{Reindl2026Pa13} suggested that its eccentricity may have been excited by the CE process. Here, we show that even with zero initial eccentricity, the disk--binary interaction can raise the eccentricity to the observed Pa~13 value within \(\sim 2500\) yr (Model~A, \(M_{\rm disk}\gtrsim5\times 10^{-5}M_\odot\)), \(\sim 400\) yr (Model~B, \(M_{\rm disk}\gtrsim 3\times 10^{-4}M_\odot\)), or \(\sim 60\) yr (Model~C, \(M_{\rm disk}\gtrsim 2\times 10^{-3}M_\odot\)) after CE ejection, offering an alternative or complementary explanation independent of CE-induced eccentricity excitation.

\section{Implications for future GW detection}
\label{sec:lisa}
DWDs are among the most important GW sources for space-borne detectors \citep{Ruiter2010,Korol2017}. Previous studies that did not consider a CBD generally assumed that DWDs as GW sources are in circular orbits. However, our results suggest that extremely short-period DWDs with a CBD may have non-negligible eccentricities. The eccentricity can affect both the GW merger timescale and the GW signal strain amplitude. Here we examine how the disk–binary interaction affects the GW signatures.

In order to characterize this influence, we compute LISA signal-to-noise ratios (SNRs) with the LEGWORK package \citep{LEGWORK_joss, LEGWORK}, which uses the instrumental and Galactic confusion noise power spectral densities from \citet{Robson_noise}. As in, e.g. \citet{Korol2017, Lambertslisa, Amaro-Seoane2023, ThieleDWDpopsynth}, we consider a DWD to be individually resolvable by LISA if its SNR exceeds 7. We model the influence of a $\tau_{\rm disk} = 10^5 \;\rm yr$ CBD on the LISA SNR using the example of an $M_1=1.25\,M_\odot$,
$M_2=0.28\,M_\odot$ DWD with a post-CE period of $P_0=0.1$ d and initial eccentricity $e=1.5\times10^{-7}$, located at a distance of 1 kpc and observed with a 4 yr LISA mission. Eccentricities are accounted for by square-summing the SNRs for the 100 first GW harmonics. We distinguish three cases: $M_{\rm disk}=0$ (referred to as ``GW-only''), $M_{\rm disk}=10^{-5}M_\odot$, and $M_{\rm disk}=10^{-3}M_\odot$. This enables a direct comparison of the LISA SNR evolution between a GW-only evolution, as implemented in most BPS studies \citep[e.g.][]{Korol2017, Lambertslisa, ThieleDWDpopsynth}, a system that merges after the disk is disrupted ($M_{\rm disk}=10^{-5}M_\odot$), and a system that merges while interacting with the disk ($M_{\rm disk}=10^{-3}M_\odot$).

\begin{figure}[!t]
\centering
\includegraphics[width=\columnwidth]{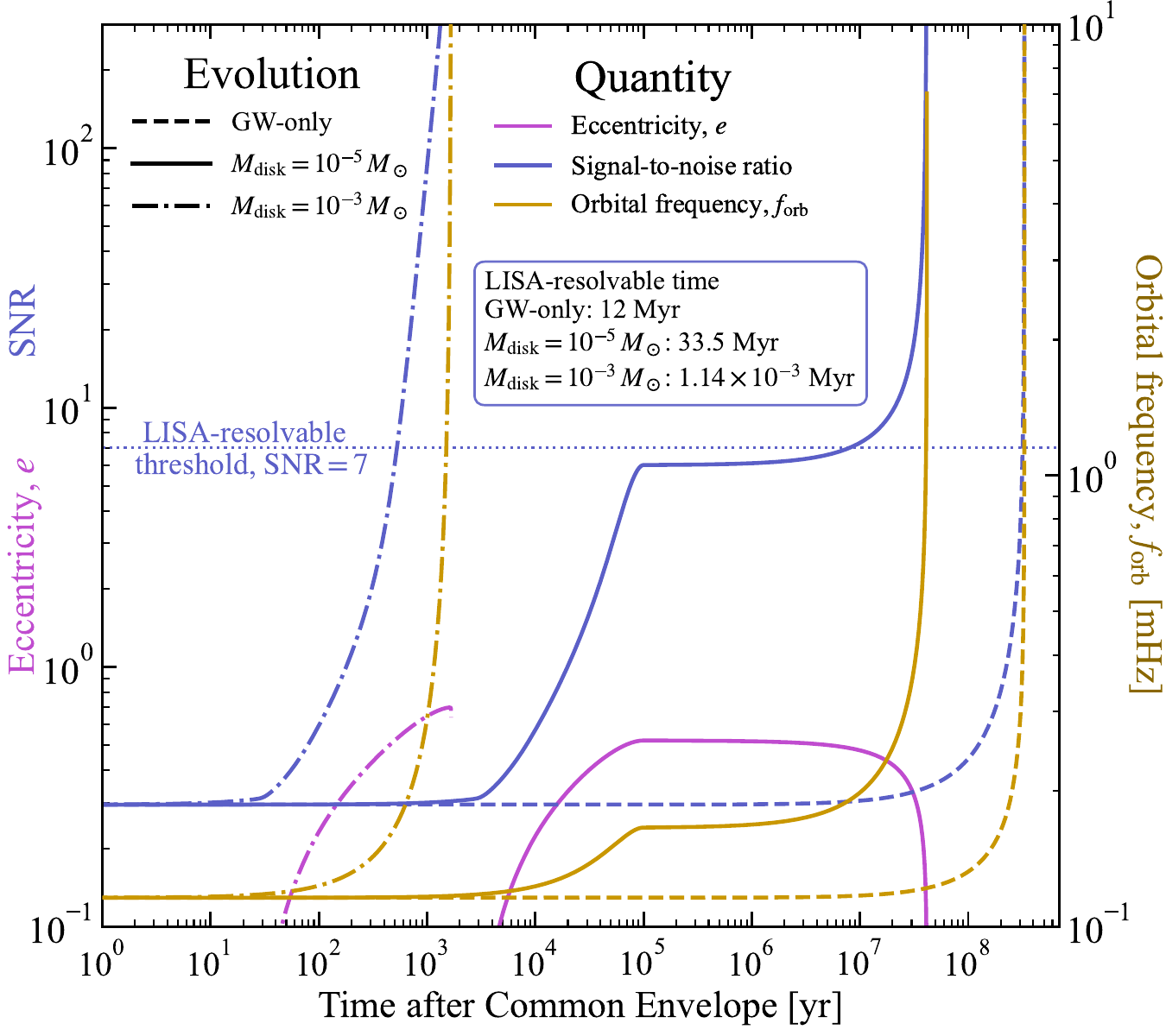}
\caption{Evolution of the eccentricity (pink), orbital frequency (gold, right-hand axis) and SNR (blue) of an $M_1=1.25\,M_\odot$, $M_2=0.28\,M_\odot$ DWD with post-CE period $P_0=0.1$ d and initial eccentricity $e=1.5\times10^{-7}$, at 1 kpc from a 4 yr LISA mission. This illustrative comparison isolates disk-mass effects: GW-only evolution (dashed lines) is compared with $M_{\rm disk}=10^{-5}M_\odot$ (solid lines) and $10^{-3}M_\odot$ (dash-dotted lines) CBD cases, both with a fixed $\tau_{\rm disk}=10^5\,{\rm yr}$ rather than the period-dependent BPS lifetimes. The LISA-resolvable threshold is set at SNR = 7 and is shown with a horizontal blue dotted line. The time spent above that threshold is shown in a text box for the GW-only, $M_{\rm disk}=10^{-5}M_\odot$ and $M_{\rm disk}=10^{-3}M_\odot$ (respectively, 12 Myr, 33.5 Myr and 1140 yr).}
\label{fig:SNR}
\end{figure}

Figure \ref{fig:SNR} shows the orbital evolution and LISA SNR for GW-only, $M_{\rm disk}=10^{-5}M_\odot$, and $M_{\rm disk}=10^{-3}M_\odot$. For comparison, we plot, as dashed lines in Figure \ref{fig:SNR}, the evolution of the system under GW radiation alone as commonly implemented in BPS \citep[e.g., ][]{Lambertslisa,ThieleDWDpopsynth}. With this prescription, the system merges after 334.14 Myr. It is resolvable by LISA for 12 Myr and its eccentricity decreases from $e=1.5\times10^{-7}$ with time.

The system with an $M_{\rm disk}=10^{-5}M_\odot$ CBD (solid lines) experiences an increase in eccentricity and orbital frequency during the lifetime of the disk, as presented in Sections \ref{sec:ecc_or_p}. At $t>10^5$yr, the disk is disrupted, and the subsequent evolution is driven solely by GW radiation before merging after 41.54 Myr (earlier than the GW-only system). This system is resolvable by LISA during 33.5 Myr, i.e. most of its lifetime, and almost three times the GW-only resolvable duration. This is because the disk excited the eccentricity of the system, saturating at $e\approx0.5$, increasing its SNR earlier, and causing it to exceed SNR = 7 after $\approx 8$ Myr instead of $\approx 200$ Myr for the GW-only system. Therefore, the system with an $M_{\rm disk}=10^{-5}M_\odot$ CBD contributes to the LISA power spectrum density for a longer time, starts earlier in its lifetime and has a fossil eccentricity $e > 0.3$ during most of its resolvable time. This system is therefore LISA-resolvable after the end of the lifetime of its disk and contains detectable fossil orbital information on its former CBD. 

On the other hand, the system with an $M_{\rm disk}=10^{-3}M_\odot$ CBD (dash-dotted lines) merges before disk depletion, 1.43 kyr after formation. Because of its very short lifetime, this binary contributes much less to the LISA confusion noise than its counterparts, and is less likely to be detected by LISA as it stays resolvable for only 1.14 kyr. If such a detection occurs regardless, the presence of a CBD could be directly identified by the eccentricity $e > 0.5$ and the faster chirping of the disk-driven merger. This system would be a clear outlier in the LISA DWD populations. 

Therefore, a low-mass CBD can enhance a DWD's contribution to the Galactic LISA-band GW foreground and population by exciting a fossil eccentricity and shrinking its orbit, whereas a sufficiently massive CBD can nearly eliminate the system's contribution through a disk-driven merger. Both of these changes may influence the GW predictions from BPS.

\section{Conclusions} 
\label{sec:conclusions}
In this work, we investigated how the disk--binary interaction affects DWD orbits, assuming that some of the CE material can form a CBD around the DWD. We found that the resonant interaction with the CBD can continuously excite the eccentricity of the DWD. The growth rate of the eccentricity increases with the disk mass and lifetime. For systems with the other parameters fixed, the resonant interaction is more efficient at short orbital periods. Based on the assumed models of disk lifetime, we also performed a BPS of DWDs with CBD interaction. In the BPS, we adopt three disk lifetime models. In Model A, the disk lifetime is proportional to the binary separation, $\tau_{\rm disk}\propto a$, in Model B, it follows the viscous scaling $\tau_{\rm disk}\propto a^{3/2}$, and in Model~C, it is proportional to the square of the binary separation, $\tau_{\rm disk}\propto a^2$.

At a fixed initial period, the disk lifetime is longest in Model~A and shortest in Model~C, with Model~B in between. The stronger excitation toward shorter periods in Model~A reflects the competition between stronger resonant forcing and shorter disk lifetimes, whereas Model~C's steeper separation dependence favors longer-period systems. To explain the observed eccentric DWDs, Model A requires a relatively small disk mass ($\sim 3$--$4\times 10^{-5}M_\odot$) and predicts that there may be DWDs with considerable eccentricities ($\gtrsim 0.1$) at very short orbital periods ($\lesssim 0.1\;\rm d$). Model B, with the viscous scaling, requires an intermediate disk mass ($\sim 3$--$3.5\times 10^{-4}M_\odot$) and predicts long-period DWDs with $e\sim10^{-4}$--$10^{-2}$. Model~C requires a massive disk ($\sim 3$--$4\times 10^{-3}M_\odot$) and predicts that there are many DWDs with eccentricities larger than 0.1 at long orbital periods ($\gtrsim 1\;\rm d$). Future observations of eccentric long-period DWDs are needed to constrain the corresponding disk parameters.

The existence of eccentric short-period systems during or after the CBD–binary interaction has important
implications for future space-borne GW detectors. For a low-mass CBD (e.g., $\sim 10^{-5}\,M_\odot$), the disk shrinks the orbit and excites eccentricity, thereby extending the LISA-resolvable phase and leaving a fossil eccentricity after disk dispersal, which offers a potential probe of past disk interaction. A more massive CBD (e.g., $\sim 10^{-3}\,M_\odot$), in contrast, drives the binary to merge before disk depletion, reducing the LISA-resolvable phase. Although such systems are less likely to be observed, their high eccentricities and rapid chirping could provide evidence of ongoing disk interaction. These contrasting outcomes suggest that CBD interaction should be accounted for in population synthesis predictions of LISA-resolvable DWDs and the Galactic GW foreground.

\begin{acknowledgments}

This study was supported by the National Key Research and Development Program of China (grant Nos. 2023YFC2206701, 2021YFA1600403, 2021YFA1600400), the National Natural Science Foundation of China (grants Nos. 12475056, 12247101, 12125303, 12525304, 12622306, 12288102, 12090040/3, 12473034, 12273105, 12073070, and 12173081), the Fundamental Research Funds for the Central Universities (grant No. lzujbky-2025-jdzx07) , the Natural Science Foundation of Gansu Province (No. 22JR5RA389, 25JRRA799), the 111 Project under (grant No. B20063), the CAS Project for Young scientists in Basic Research (YSBR-148), the Strategic Priority Research Program of the Chinese Academy of Sciences (grant Nos. XDB1160303, XDB1160201, XDB1160000), the CAS ``Light of West China", the Yunnan Revitalization Talent Support Program-Science $\&$ Technology Champion Project (No. 202305AB350003) and Young Talent project, the International Centre of Supernovae (ICESUN), Yunnan Key Laboratory of Supernova Research (Nos. 202302AN360001, 202201BC070003), Yunnan Fundamental Research Projects (No. 202401AT070139), the Natural Science Foundation of Henan Province (No. 242300420944). Dylan Hebrail and Ross Church were supported by grant 2024-00123 - Algorithm development for LISA from the Swedish National Space Agency. {X.C. acknowledges the New Cornerstone Science Foundation through the XPLORER PRIZE.} The authors gratefully acknowledge the “PHOENIX Supercomputing Platform” jointly operated by the Binary Population Synthesis Group and the Stellar Astrophysics Group at Yunnan Observatories, Chinese Academy of Sciences. 
  \end{acknowledgments}



\bibliography{sample701}{}
\bibliographystyle{aasjournalv7}

\end{document}